# arXiv@25: Key findings of a user survey

Oya Y. Rieger
Cornell University Library
Ithaca, NY
oyr1@cornell.edu

Gail Steinhart
Cornell University Library
Ithaca, NY
gss1@cornell.edu

Deborah Cooper
Cornell University Library
Ithaca, NY
dsc255@cornell.edu

**ABSTRACT**

As part of its 25th anniversary vision-setting process, the arXiv team at Cornell University Library conducted a user survey in April 2016 to seek input from the global user community about arXiv's current services and future directions. We were heartened to receive 36,000 responses from 127 countries, representing arXiv's diverse, global community. The prevailing message is that users are happy with the service as it currently stands, with 95% of survey respondents indicating they are very satisfied or satisfied with arXiv. Furthermore, 72% of respondents indicated that arXiv should continue to focus on its main purpose, which is to quickly make available scientific papers, and this will be enough to sustain the value of arXiv in the future. This theme was pervasively reflected in the open text comments; a significant number of respondents suggested remaining focused on the core mission and enabling arXiv's partners and related service providers to continue to build new services and innovations on top of arXiv.

**1. INTRODUCTION AND BACKGROUND**

arXiv is a moderated scholarly communication forum informed and guided by scientists and the scientific cultures it serves. Established at Los Alamos National Laboratory by physicist Paul Ginsparg in 1991 as a means for researchers in theoretical high-energy physics to share their work in advance of publication, arXiv now plays a central role in the broader range of disciplines it serves, and is a model for the potential for transforming scholarly communication in general (Ginsparg 2011). Cornell University Library assumed management responsibility for arXiv when it moved with Ginsparg to Cornell in 2001. The site is now collaboratively governed and supported by the research communities and institutions that benefit from it most directly, ensuring a transparent and sustainable resource (Rieger 2011). As arXiv has grown, it has developed a business model to sustain it (Rieger and Warner 2010), with current financial support coming from the Simons Foundation, Cornell University Library, and about 190 member libraries from all around the world.

As part of its 25th anniversary vision-setting process[1], and in recognition of the critical need to modernize its infrastructure, the arXiv team at Cornell University Library (CUL) conducted a user survey in April 2016 to seek input from the global user community about the current services and future directions. This paper reports the results of that survey, and possible next steps.

**2. METHODS**

In preparation for surveying arXiv users, the arXiv team conducted a literature review and surveyed members of its Scientific and Member Advisory Boards (SAB and MAB, respectively).[2]

The literature review laid the groundwork for the survey design as the papers raised many relevant questions and brought up common themes. Some papers broadly looked at digital repositories and perceptions of self-archiving and depositing (Kim 2010; Nicholas *et al.*, 2012); others considered in detail the specifics of publishing in an open access environment and raised issues such as author fees and citation (Fowler 2011). Frequently mentioned themes included: reasons to deposit into arXiv (Nicholas *et al*., 2012) and the value of "early dissemination of research findings" (Fowler 2011); scholarly reputation; long-term preservation; adhering to standard practice in the field; the willingness to cite a preprint only (Fry *et al*., 2015), whether to publish in the open access realm specifically, and the impact on retaining rights for published papers (Fowler 2011).

Papers that addressed the effectiveness of specific features within repositories were most relevant for informing our survey design. These focused predominantly on methods of searching, for example, by known articles, subject-based search, author search and full text searching. The ease of the submission process and citation features were also commonly discussed. The literature review also uncovered specific enhancements users would like to see, such as mobile access, personalization and collaboration tools (Gentil-Beccot *et al.*, 2009).

---

[1] https://confluence.cornell.edu/display/culpublic/arXiv+Review+Strategy

[2] For information about arXiv's business and governance model, see: https://confluence.cornell.edu/x/xKSTBw

The MAB and SAB survey focused on the areas of quality control and rapid dissemination, subject area expansion, developing new services and improving on current services, and the future of arXiv. The survey of the MAB and SAB then served as the basis for drafting the user survey. Both boards and arXiv staff reviewed and helped refine the final instrument. Prior to deployment, the survey instrument was tested by a group of users representing scientists and students. CUL staff prepared and deployed both the preliminary survey of SAB and MAB members as well as the user survey in Qualtrics, a web-based survey tool. The survey was deployed for three weeks (April 6-26, 2016), and publicized via a banner on the arXiv.org website, various email lists, and social media. CUL has a blanket exemption from Cornell's Institutional Review Board that allows its staff to conduct research that is focused on its products and services (i.e. not human participant research). For readability of figures in the main body of this paper, we have shortened some of the questions; the complete survey instrument is included in Appendix 1.

The survey consisted of 17 multiple-choice and 8 open-ended questions. For the multiple-choice questions, quantitative data analysis was focused on descriptive statistics. Analysts examined differences in response type (levels of importance, preferences, etc.) by demographic groupings and by years of arXiv use. The large sample size resulted in overall statistically significant results for all differences we tested, and we found it more meaningful to focus on substantial percentage differences in responses, rather than statistically significant differences.

For the eight open-ended questions, the number of responses per question ranged from 853-3374. Each question was assigned to an analyst. Given the volume of responses, analysts read through a random sample of 10-20% of the responses for each question to develop codes for response topics. Once codes were established for each question, the analysts read or re-read carefully at least 20% of the responses, coding the responses in their sample, and noting positive or negative sentiments as appropriate. Analysts then skimmed all remaining responses and noted topics and trends not present in the sample. Some analysts used http://voyant-tools.org/ to generate word clouds for a visual representation of word frequencies in the text responses. For each question, the corresponding analyst noted recurring themes and ordered them according to frequency of mention in open-ended responses, and produced a short summary report of their findings.

## 3. RESULTS AND DISCUSSION

We were heartened to receive 36,000 responses (28,233 fully completed) from 127 countries, representing arXiv's diverse community (complete survey results, including demographic information, in Appendix 2). Between 2.4-9.4% of the total respondents answered each open-ended question. While this is a relatively small proportion of respondents, the quantity of input we received in their answers is substantial, with surprising consistency in the comments. The prevailing message is that users are happy with the service as it currently stands. 95% of survey respondents said that they are very satisfied or satisfied with arXiv. Furthermore, 72% of respondents indicated that arXiv should continue to focus on its main purpose, which is to quickly make available scientific papers, saying this will be enough to sustain the value of arXiv in the future. This theme was pervasively reflected in the open text comments. A significant number of respondents suggested keeping to the core mission (rapid dissemination and preservation, open access, community moderation of scholarly articles) as well as enabling arXiv's partners and related service providers to continue to build new services and innovations on top of arXiv. That said, respondents did note areas for improvement or development. Below, we summarize the findings for each of the survey topics (current services, potential new services, finding arXiv papers, quality control and rapid dissemination, subject area expansion, and the future of arXiv).

### 3.1 Improving current arXiv services

When asked about the importance of improving a specific range of services, more than 70% of respondents said that improving search functions to allow more refined results was *very important/important* across all groups by years of use, age groups, number of articles published, country groups, and subject areas (Figure 1). Many commenters requested enhanced functions such as author search, date-limited searching, and searching non-English languages. Search was equally problematic regardless of whether the user searched for a known paper, was browsing a subject category, or looking for specific authors.

A series of questions asked users about improving the submission process specifically with (1) support for submitting research data, code, slides and other materials; (2) improving support for linking research data, code, slides, etc., with a paper; and (3) updating the TeX engine and various other enhancements (Figure 1). For most questions, about 40% of respondents rated each one as *very important/important*. The notable exception is search: respondents overwhelmingly favor improving arXiv's search functionality.

The open text responses demonstrated considerable interest in better support for supplemental materials, although responses were divided as to whether they should be hosted by arXiv or another party. Many respondents were supportive of integrating or linking to other services (especially GitHub), while a significant number of respondents also indicated doubts about long-term availability and link rot for content not hosted within arXiv. Some expressed concerns regarding the resources required for arXiv to improve this. There was some interest in including the data underlying figures in arXiv papers. Among other services and improvements recommended by respondents were:

- Consistent inclusion of information and links about the published versions of the papers.
- More refined options for alerting, both email and RSS. Several respondents specifically requested email alerts for works by a particular author, and there was some interest in HTML-formatted email with live links.
- Updating arXiv's TeX engine and providing TeX templates or style files to make submission easier.
- Linking papers to each other via citations and actionable links in bibliographies.
- Ability to submit a PDF, an increase in the file size limit (often with the specific request to link to figures), and the ability to upload multiple files at once.
- Allowing submission directly from authoring platforms (such as Overleaf or Authorea).
- Providing use statistics such as paper downloads and views.

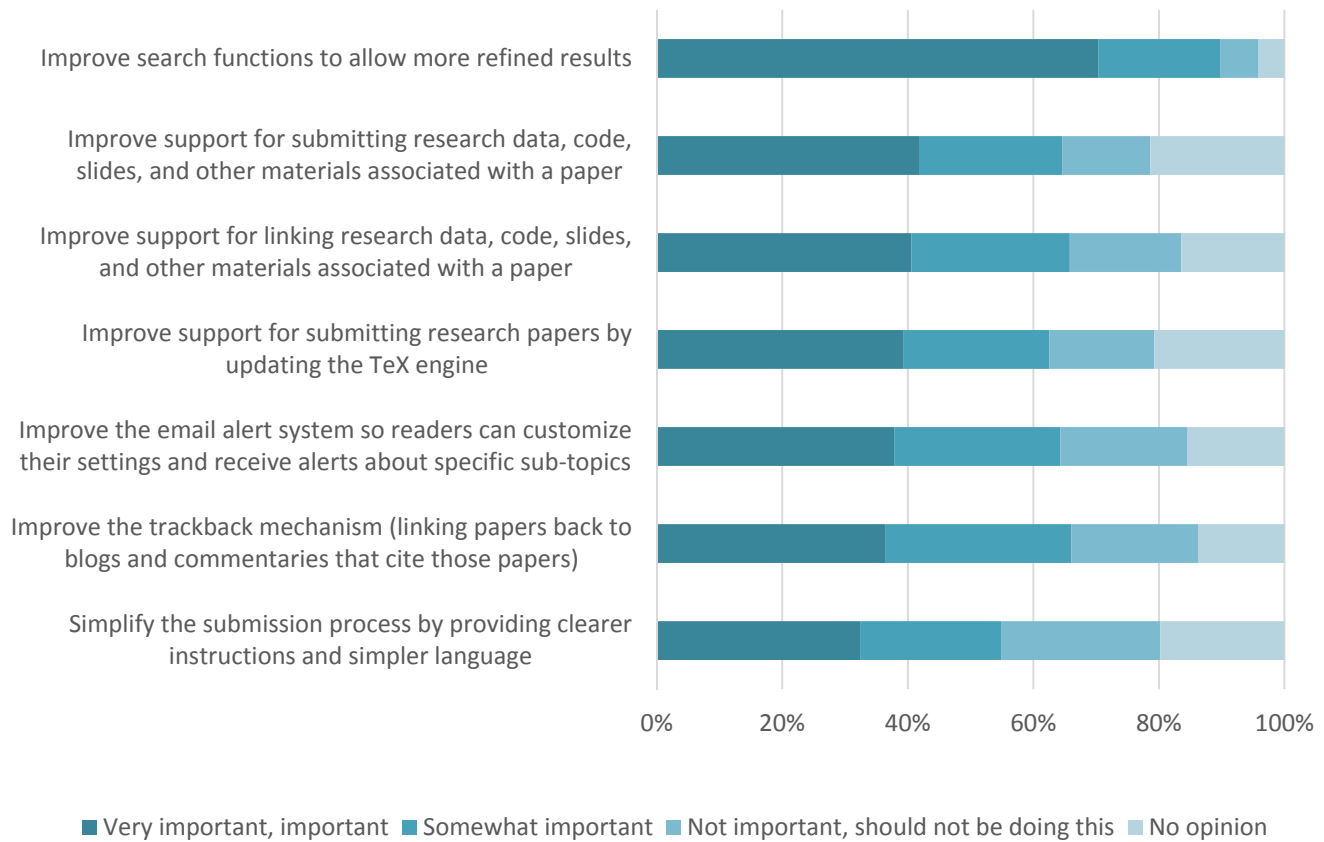

Figure 1. Importance of improving selected current services.

Interestingly, regarding improvements to current services, more recent arXiv users (five years or less) selected the "no opinion" option more than experienced users. For example, Figure 2 shows the responses for the question "Improve support for submitting research papers by updating the TeX engine." For all the questions in this category, the same trend is evident: a higher percentage of relatively new users expressed that they had no opinion and this percentage of respondents decreased with each level of increase in years of use. Interestingly, this same trend is not visible by age group; i.e., our data do not show that a higher percent of younger users have no opinion on improvements to current services.

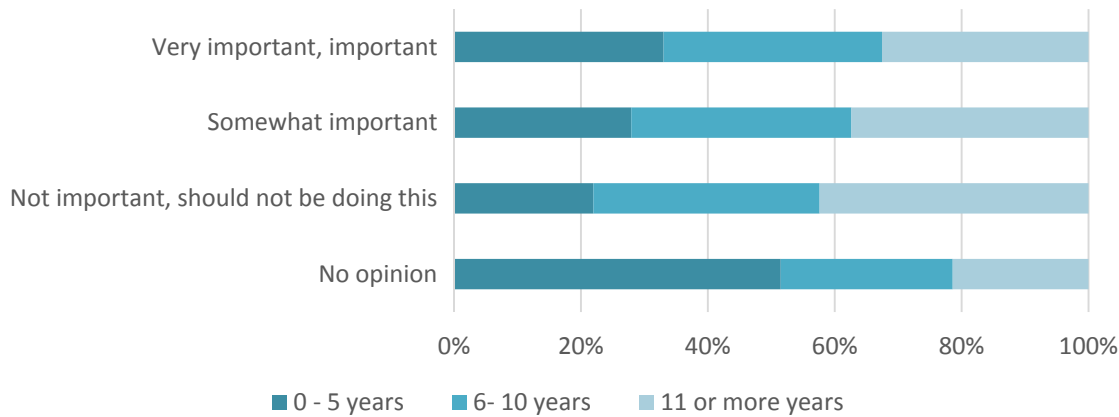

Figure 2. Responses to the question "How important is it to… Improve support for submitting research papers by updating the TeX engine," by years respondents have used arXiv.

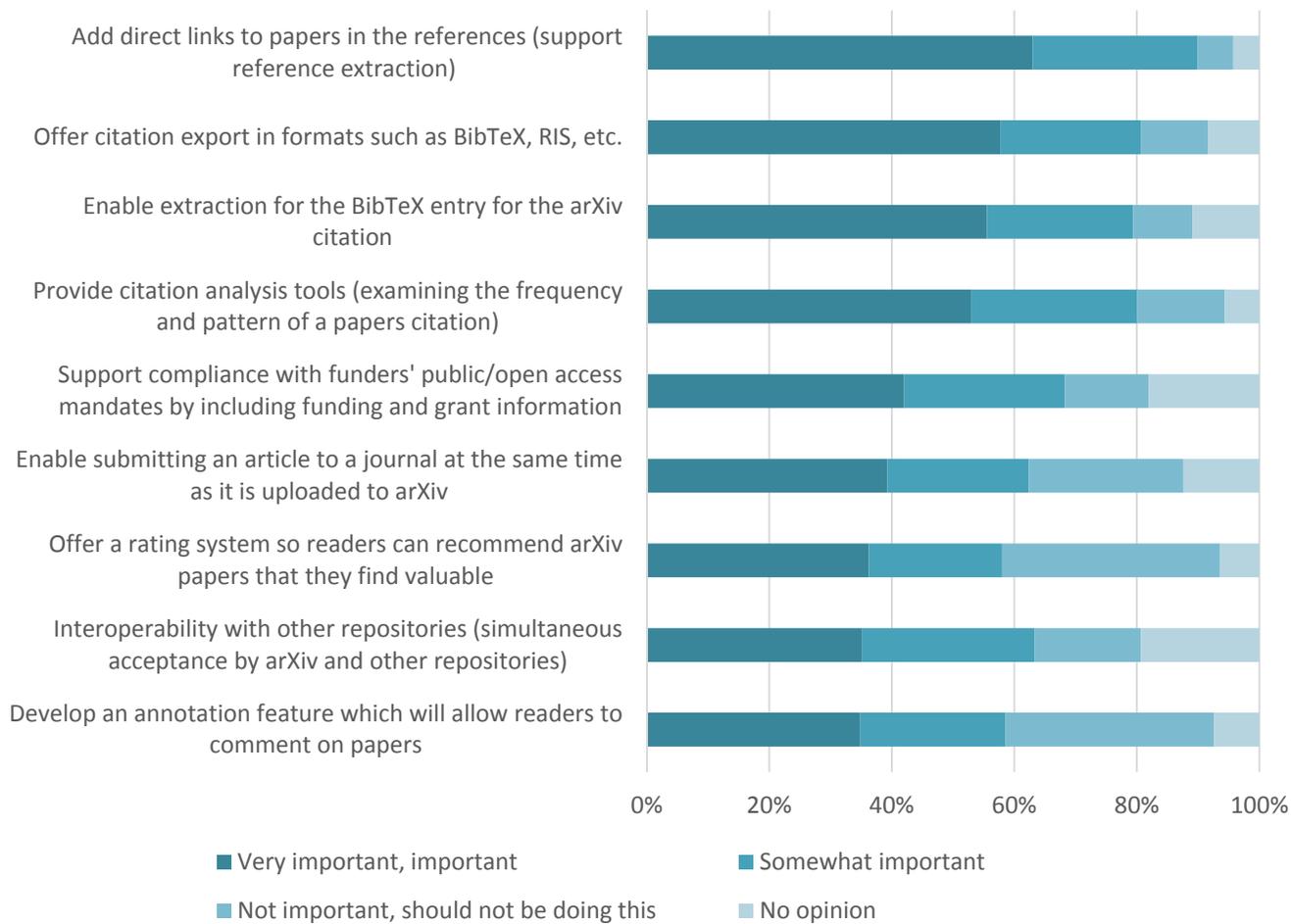

Figure 3. Importance of developing new arXiv services.

## 3.2 Developing new services

Users were asked to rate a range of proposed new services for arXiv (Figure 3). In the ranked responses, more than 63% of users rated adding direct links to papers in the references (reference extraction) *as very important/important*. Citation export in formats such as BibTeX, RIS was rated as *very important/important* by more than 57% of users, and extraction for the BibTeX entry for the arXiv citation was similarly rated by more than 55% of respondents. Citation analysis tools in general were ranked as *very important/important* by almost 53% of respondents. In the comments, opinions were divided on the need for enhanced citation-analysis capabilities. While users were generally in favor of citation tools many of the same users noted that other systems are already doing this, and that this was sufficient for their needs.

Responses to the question about offering "a rating system so readers can recommend arXiv papers that they find valuable" were closely split between very important/important (36%) and not important/should not be doing this (36%). However, recent users (0-5 years) favored a rating/recommendation system more strongly (42%) than did seasoned ones (34% for 6-10 years and 28% for 11 or more years), and a larger percentage of younger users considered it important (42% of those under 30 years), as compared to 28% of those 60-69 and 70 and above (Figure 4a, b). Opinions were divided in the open text comments but overall, respondents were hesitant about the idea. Some users liked the rating feature "in an ideal world" setting, but did not think it was appropriate for arXiv; others expressed concern that it would dilute the mission of arXiv, or that it simply appears infeasible in the current version of arXiv. Even users in favor of a rating system point to the complexity of implementing such a system, raising issues as to whether it would be completely open to the public, open only to peers, anonymous, etc. Several respondents stressed that such a feature would need to be implemented very carefully.

Like the question about offering a rating system, the idea of adding an annotation feature to allow readers to comment on papers was almost evenly split, with 35% of users ranking it as *very important/important* and 34% as *not important/should not be doing this*. In the open text responses, the trend was opposition to the idea and some of the responses reflected strongly negative opinions. Those in favor or open to the idea of a commenting system often added a caveat, and in general there was a sense of caution even for those responding positively. A common concern was that a moderated system and verifiable accounts would be necessary to prevent a free-for-all. Unlike the question about offering a rating system, there were no discernible differences in opinion based on different demographic characteristics.

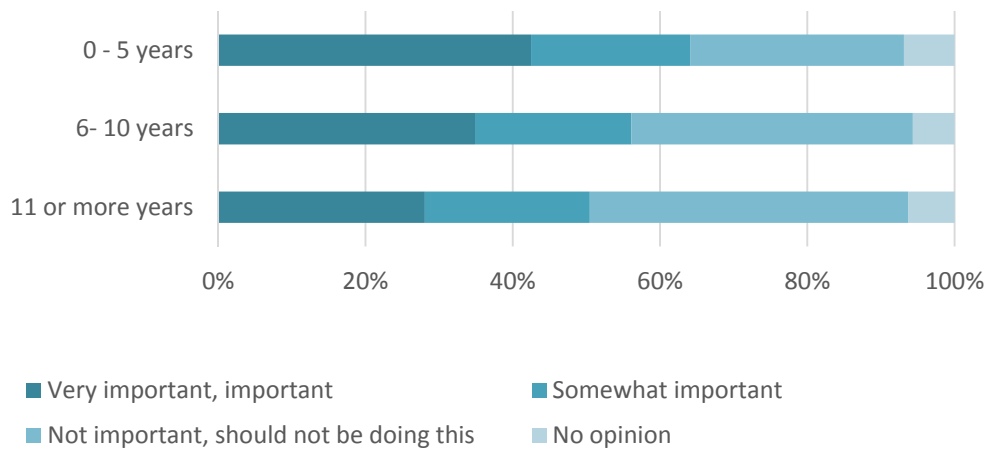

Figure 4a. Responses to the question "How important is it to… Offer a rating system so readers can recommend arXiv papers that they find valuable," by years respondents have used arXiv.

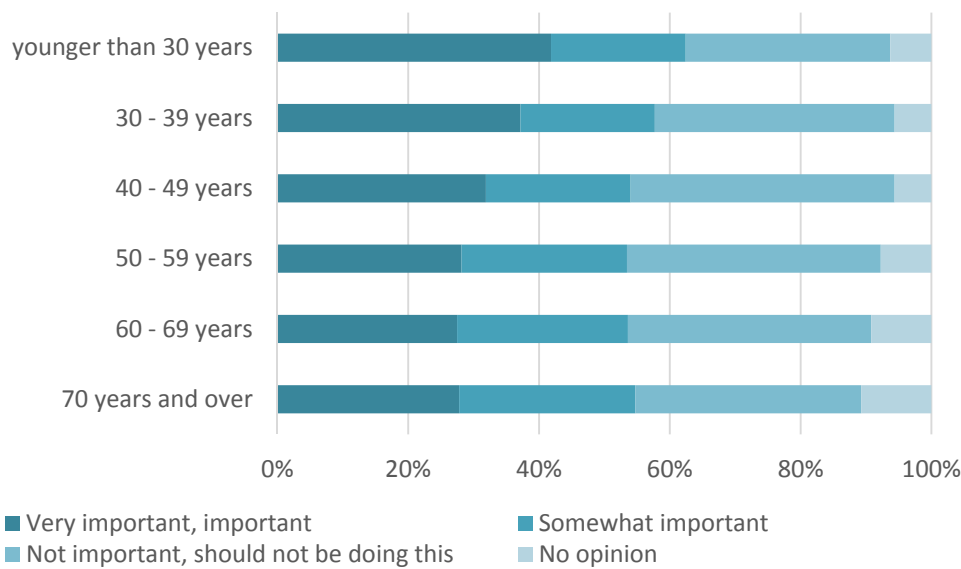

Figure 4b. Responses to the question "How important is it to… Offer a rating system so readers can recommend arXiv papers that they find valuable," by respondent age.

**3.3 Finding arXiv papers**

The vast majority of arXiv's users access the papers directly from the homepage (79%), followed by using Google to search (50%) and Google Scholar (35%) (Figure 5a). Once on the homepage, reactions were mixed regarding the ease of use and navigation. To discover content, 63% of users *go to the link for new or recent under a particular category* and equally 63% of users *use arXiv's search engine and enter a specific arXiv ID, author name or search term.* A small number of users, 14%, rely on the daily mailing list and then look for a particular article in the search field (Figure 5b). User ratings vary regarding ease of use of the arXiv home page for finding papers: 32% rated it as *easy,* but only 25% find it *somewhat easy* and 21.6% rated it *somewhat difficult* to use (Figure 5c)

In the open text comments, opinion was divided about the user interface. The majority of respondents disliked the outdated style, but a vocal minority appreciated the interface's simplicity, which these users feel helps arXiv efficiently focus on its core mission. Though some users suggested new or additional features, a majority of respondents emphasized that the clean, uncluttered nature of the site makes its use easier and more efficient. "I sincerely wish academic journals could try to emulate the cleanness, convenience, and user-friendly nature of the arXiv, and I hope the future of academic publishing looks more like what's we've been able to enjoy in the arXiv," one user wrote. Additional issues mentioned were the number of links, layout and finding submission information. The lack of hierarchical organization was noted as a particular challenge to understanding arXiv's navigation.

Requests for enhancements related to the user experience included greater personalization of arXiv for readers; for example, the ability to "favorite" papers, curate a personal library, and see recommendations when users visit the site. Other users mentioned the development of APIs to further facilitate the development of overlay journals. Some users also suggested the development of a mobile-friendly interface.

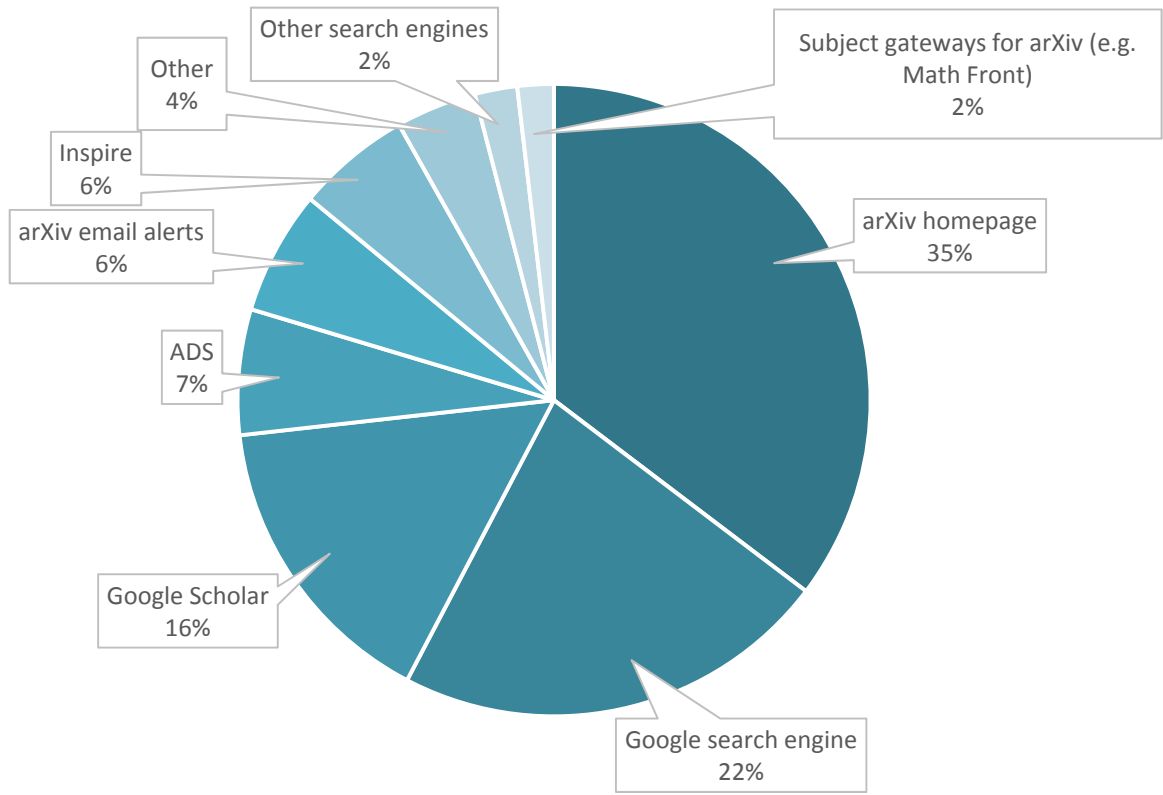

Figure 5a. Responses to the question, "Where do you go to find arXiv papers?"

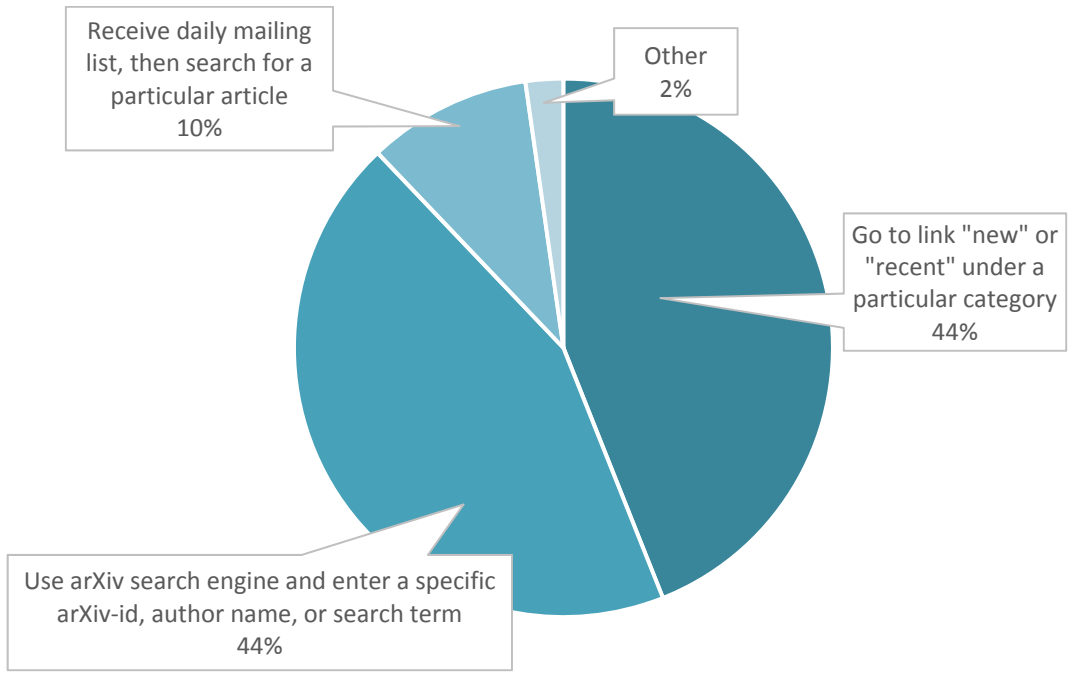

Figure 5b. Responses to the question, "If you have used the arXiv homepage, how do you usually navigate our main page?"

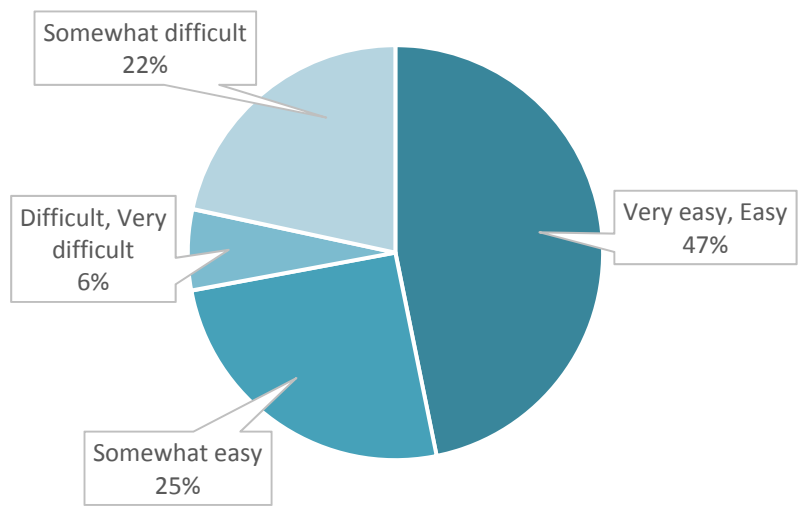

Figure 5c. Responses to the question, "If you have used the arXiv homepage for finding papers, how easy is it to navigate?"

Many commenters either described how they rely on other services to interact with arXiv content (site-specific search engine searches, the SAO/NASA Astrophysics Data System, ADS, and INSPIRE, the High-Energy Physics Literature Database) or recommended features based on their experience with other information systems. Among those frequently praised were ADS, INSPIRE, Google Scholar, gitxiv.com and arxiv-sanity.com.

**3.4 Importance of quality control**

26,430 arXiv users responded to a series of questions regarding quality-control measures (Figure 6). The most important of these (ranked very important/important) were: Check papers for text overlap, i.e., plagiarism (77%), Make sure submissions are correctly classified (64%), Reject papers with no scientific value (60%) and Reject papers with self-plagiarism (58%).

There were no discernible differences across demographic groups for all quality control measures. Self-plagiarism was mentioned as an area for refinement, with some users noting that context is key. For example, conference papers are a common and typical area where self-plagiarism could legitimately occur in a scientifically sound submission.

Several respondents said they were unaware of precisely what quality-control measures were already in place, and felt that the process is too opaque. Others acknowledged the difficult balance between rejecting papers that are clearly unworthy—"crackpot"—and rejecting papers for other, perhaps less obvious or transparent reasons. However, even in the face of such criticisms there was a strong thread of satisfaction with arXiv's current quality-control process and users cautioned against going too far in the other direction.

Some users would prefer that arXiv embrace a more open peer review and/or moderation process, while others were adamant that current controls allow arXiv the freedom and speed of access that is otherwise unobtainable through traditional publishing.

Overall, the feeling was that quality control matters but user comments varied greatly in relation to how arXiv could practically achieve these goals. As one respondent wrote, "Judgment about quality control is a very relative issue."

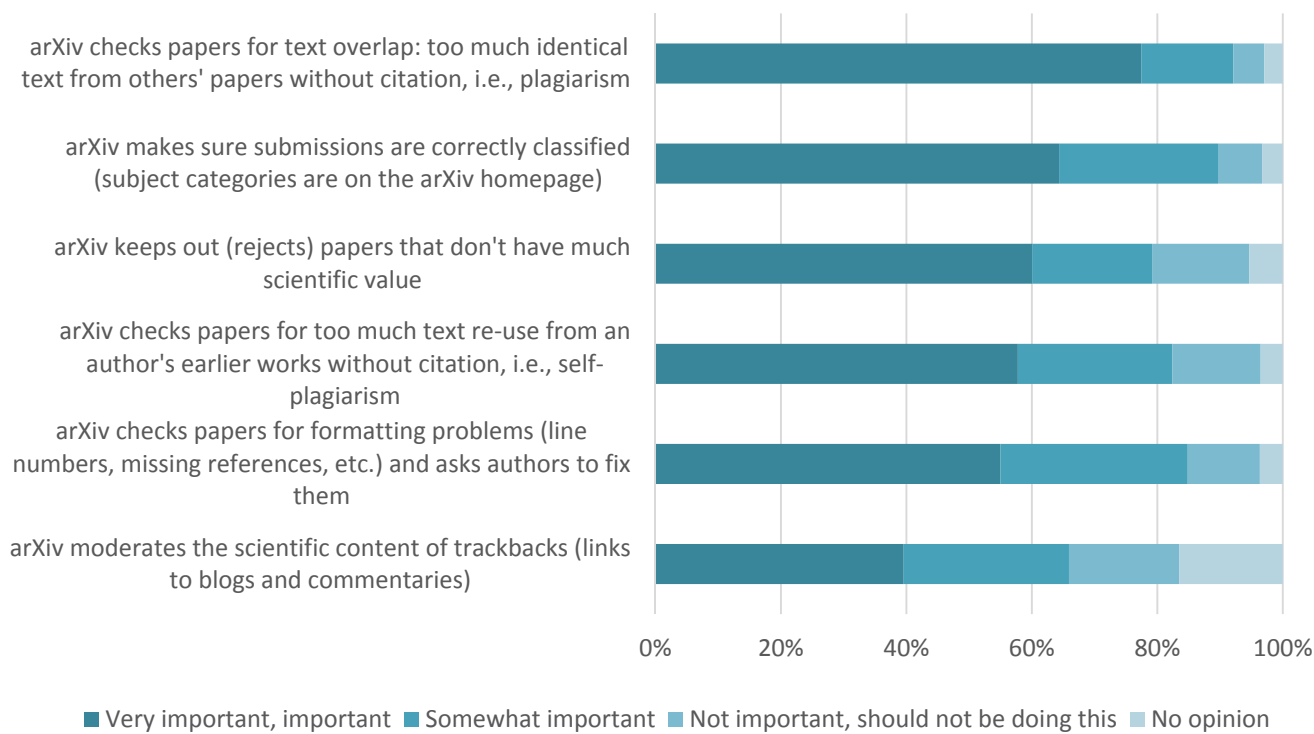

Figure 6. Responses to the question "How important are the following CURRENT quality control measures?"

### 3.5 Subject area expansion

New subject categories are generally not a priority for arXiv users. 73% of the respondents were not interested in seeing new subject categories added to arXiv, although 26% of respondents would like to see new subject categories added and suggested Chemistry (881), Engineering (483), Biology (429), Economics (248), Philosophy (220), and Social Sciences (106). There were also several more specialized categories suggested, such as Machine Learning (82 responses) and Artificial Intelligence (27 responses).

A frequently repeated theme was that arXiv does not need to focus particularly on additional subjects but instead should focus on the refinement and addition of subfields and subcategories, especially in High Energy Physics Theory as well as Mathematics.

### 3. 6 Overall perceptions of arXiv

Many of the comments reflected deep satisfaction with and gratitude for arXiv. Several users referred to the significance of the service for their personal career development and expressed thanks for its continued existence; for example, a typical comment was: "Thanks for the hard work of many people over the years. My work life would be very different without your efforts." arXiv also received many plaudits for advancing the dissemination of research through the open-access system. One user referred to the service as "a beacon for scientific communication." Several commenters expressed how crucial arXiv has been for them personally in enabling them to quickly access the latest research in their field. There was an overall perception that arXiv was an important leader in the development of alternatives to traditional publishing. Independent researchers who are unaffiliated with large institutions and who might otherwise have delayed access to papers particularly emphasized the importance of arXiv for their work.

The combination of multiple choice responses (Appendix 1 for the complete survey instrument) and the extensive and thoughtful open text comments pinpointed areas that need to be upgraded and enhanced. Improving the search function emerged as a top priority as the users expressed a great deal of frustration with the limited search capabilities currently available, especially in author searches. Providing better support for *submitting* and *linking* research data, code, slides and other materials associated with papers emerged as another important service to expand. Regardless of their subject area, users were in agreement about the importance of continuing to implement quality control measures, such as checking for text overlap, correct classification of submissions, rejection of papers without much scientific value, and asking authors to fix format-related problems.

Many users commented on the need to randomize the order of new papers in announcements and mailings. There were several useful remarks about the need to improve the endorsement system and provide more information about the moderation process and policies.

In regard to arXiv's role in scientific publishing, some users encouraged the arXiv team to think boldly and further advance open access (and new forms of publishing) by adding features such as peer review and encouraging overlay journals. On the other hand, many users strongly emphasized the importance of sticking to the main mission and not getting side-tracked with formal publishing. There was a similar divergence of opinion about encouraging an open review process by adding rating and annotation features. When it comes to adding new features to arXiv to facilitate open science, the prevailing opinion was that any such features need to be implemented very carefully and systematically, and without jeopardizing arXiv's core values.

While many respondents took the time to suggest future enhancements or the finessing of current services, several users were strident in their opposition to any changes. Throughout all of the suggestions and regardless of the topic, commenters unanimously urged vigilance when approaching any changes and recommended caution in integrating social media features into arXiv. One respondent wrote: "Do not make arXiv into a social-media platform or something complicated. Keep working on improving your core, which is what we use and love!" The feeling is that arXiv as it exists is working well and while there are some areas for improvement, too much change could potentially weaken the effectiveness and overall mission of arXiv.

**4. NEXT STEPS FOR arXiv**

arXiv's core infrastructure is fragile, running on an assemblage of legacy code written for a much smaller system. With that in mind, and given the importance of the system to the communities which it serves, arXiv is in great need of an infrastructure upgrade (Van Noorden 2016). Soliciting feedback from users is but one step in the process of developing a vision for a robust and modern arXiv.org.[3] The arXiv team plans one more survey of a key stakeholder group: its volunteer moderators. The findings of this survey will help refine immediate IT development priorities going forward, as well as the larger effort to modernize arXiv's infrastructure. In addition to soliciting in-depth feedback from users and moderators, arXiv held an IT infrastructure workshop April 26-28, 2016.[4] Participants considered a range of approaches to arXiv's future architecture, as well as specific technologies and possible sources of funding. The arXiv team is now exploring possible partnerships and funding sources, and envisions a robust future for arXiv.

ACKOWLEDGEMENTS

Many individuals were involved in designing and testing the survey and helped out with the data analysis. Special thanks go to Andrea Salguero-Cruz, Jim Entwood, Martin Lessmeister, Leah McEwen, Chloe McLaren, Chris Myers, David Ruddy, Vandana Shah, Simeon Warner, and Jake Weiskoff. Also, we are grateful for the guidance from the arXiv's Member Advisory Board and Scientific Advisory Board for their early input into the design of the survey.

---

[3] https://confluence.cornell.edu/display/culpublic/arXiv+Review+Strategy

[4] https://confluence.cornell.edu/display/culpublic/arXiv+IT+Workshop

# Appendix 1. Survey instrument

Dear arXiv User,

As an open-access site, arXiv serves people like you all over the world and your opinion counts. Please complete this questionnaire to help us improve arXiv and think of future directions for the service in a way that best serves users like you. The survey has four sections, and will take about 10 minutes to complete. We do not collect any information that can identify you and we will share only a summary of the results.

Thank you!
Cornell University Library arXiv Team

**SECTION 1**

Please tell us about yourself, so we can understand the needs of our different types of users.

I use arXiv in the following ways: (Please choose all that apply)
- I am an arXiv reader
- I am an arXiv author
- I am an arXiv submitter
- I am an arXiv (other type of user): Please describe __________

The number of articles I have published/submitted on arXiv is:
- 1 article
- 2 articles
- 3-4 articles
- 5-10 articles
- More than 10 articles

As a user, my main subject area of interest in arXiv is: (please choose all that apply):
- Physics
- Mathematics
- Computer Science
- Quantitative Biology
- Quantitative Finance
- Statistics
- Other (please specify): __________

Would you like to see additional subject categories added to arXiv?
- No
- Yes

Which subject categories would you like to see added to arXiv? (open-ended response)

I have been using arXiv for:
- 0-2 years
- 3-5 years
- 6-10 years
- 11 or more years

My current occupation is: (Please choose ALL that apply):
- I am an academic faculty member (professor) at a college or university
- I am an academic staff member (researcher or postdoc) at a college or university
- I am a researcher at a non-profit or governmental agency

- I am a Masters/Ph.D. student
- I am an undergraduate student
- I am (please describe): __________

My age is:
- younger than 30 years
- 30-39 years
- 40-49 years
- 50-59 years
- 60-69 years
- 70 years and over

My main place of work is located in: __________

Thank you for telling us about yourself. In the next section, we'll ask you about your opinion on specific areas of arXiv.

## SECTION 2

**Improving on Current Services and Developing New Services**

We welcome your suggestions and comments in the text boxes that follow the questions in each section.

**How important is it to improve on the following CURRENT arXiv services?**
(Very important, Important, Somewhat important, Not important, Should not be doing this, No opinion)
- Simplify the submission process by providing clearer instructions and simpler language.
- Improve support for submitting research papers by updating the TeX engine.
- Improve support for submitting research data, code, slides, and other materials associated with a paper (e.g., I want to be able to upload my datasets/machine readable tables with my article).
- Improve support for linking research data, code, slides, and other materials associated with a paper (e.g., I want to be able to link to my slides on SlideShare).
- Improve search functions to allow more refined results (e.g., narrow down results by additional search terms, filter by publication year or institutional affiliation, etc.).
- Improve the email alert system so that readers can customize their settings and choose to receive alerts about specific sub-topics.
- Improve the trackback mechanism (linking papers back to blogs and commentaries that cite those papers).

**Do you have suggestions for any of the above-mentioned current services?** (open-ended question)

**How important is it to develop the following NEW arXiv services?**
(Very important, Important, Somewhat important, Not important, Should not be doing this, No opinion)
- Enable submitting an article to a journal at the same time as it is uploaded to arXiv.
- Provide Citation Analysis tools (examining the frequency and pattern of a paper's citation).
- Offer citation export in formats such as BibTeX, RIS, etc.
- Enable extraction for the BibTeX entry for the arXiv citation.
- Add direct links to papers in the references (support reference extraction).
- Support compliance with public/open access mandates (funding agency policies that require research results to be made public) by allowing final versions of papers to be submitted with information such as funding sources and grant numbers.
- Enable linkages (interoperability) with other repositories (e.g., run by libraries), so that a paper accepted by arXiv is accepted at the same time by the other repositories.
- Develop an annotation feature, which will allow readers to comment on papers.
- Offer a rating system so readers can recommend arXiv papers that they find valuable.

**Do you have suggestions for any of the above-mentioned new services, or any other new services you would like to see in arXiv?** (open-ended question)

**Where do you go to find arXiv papers? Please choose all that apply**.
- Go directly to arXiv.org (arXiv homepage)
- ADS
- INSPIRE
- Google scholar
- Google search engine
- Other search engines
- Subject gateways for arXiv, such as the Math Front
- arXiv email alerts
- Other (please specify): __________

**If you have used the arXiv homepage for finding papers, how easy is it to navigate?**
- Very easy
- Easy
- Somewhat easy
- Somewhat difficult
- Difficult
- Very Difficult

**Please tell us why you think the arXiv homepage is difficult to navigate**. (open-ended question)

**If you have used the arXiv homepage, how do you usually navigate our main page? Please choose all that apply.**
- Go to link "new" or "recent" under a particular category.
- Use arXiv search engine and enter a specific arXiv-id, author name, or search term.
- Receive daily mailing list, and then look for a particular article on the search field.
- Other, please explain: __________

**Do you have any additional comments on the arXiv homepage?** (open-ended question)

**SECTION 3**

**Controlling Quality of Papers in arXiv Through Moderation**

**How important are the following CURRENT quality control measures?**
(Very important, Important, Somewhat important, Not important, Should not be doing this, No opinion)
- arXiv keeps out (rejects) papers that don't have much scientific value.
- arXiv checks papers for text overlap: an author's use of too much identical text from other authors' papers, without making it clear that the text is not their own material, i.e., "plagiarism".
- arXiv checks papers for too much text re-use from an author's earlier works, i.e., "self-plagiarism" (reuse of identical content from one's own published work without citing).
- arXiv checks papers for format-related problems (line numbers in text, missing references, oversize submissions, etc.) and asks authors to fix them before they are announced.
- arXiv makes sure submissions are correctly classified (the subject categories are included on the arXiv homepage).
- arXiv moderates the scientific content of trackback (links to blogs and commentaries) before permitting the link to be added.

**Please choose any ONE of the following statements that you agree with the most:**
- arXiv should not perform quality control at all; arXiv should focus on quickly publishing papers.
- arXiv should focus on a few parts of quality control, but mostly focus on quickly publishing papers.
- arXiv needs to keep up quality control in many areas and, if needed, should take a little more time to publish papers so that quality control can be kept.
- arXiv should quickly publish and at the same time, keep up with quality control, even if that takes time away from other new activities.
- arXiv should maintain its current quality control level.

**Do you have any specific suggestions or comments about quality control in arXiv?** (open-ended question)

**SECTION 4**

**The Future of arXiv**

**Which of the following BEST describes your opinion of how arXiv needs to move forward?**

- arXiv should focus on its main purpose, which is to quickly make available scientific papers. This will be enough to hold up the value of arXiv in the future.
- arXiv should expand its main mission, and spend more time and resources to provide new services. This is necessary to hold up the value of arXiv in the future.
- No opinion.

**Overall, how satisfied are you with arXiv?**
- Very satisfied
- Satisfied
- Somewhat satisfied
- Somewhat dissatisfied
- Very dissatisfied
- No opinion

**Do you have any additional comments or suggestions on how arXiv can be improved to better meet the user's needs?** (open-ended question)

**May we contact you in case we want to respond to any of your comments? Please give us your name and email address if we have your permission to contact you.**
Name __________
Email address __________

If you have any questions about this survey, please contact support@arxiv.org.

Thank you for taking the time to answer this survey.

# Appendix 2. Complete survey results

### Section 1

I use arXiv in the following ways: (Please choose all that apply)

| Answer | % | Count |
|---|---|---|
| I am an arXiv reader | 92.83% | 31862 |
| I am an arXiv author | 53.23% | 18270 |
| I am an arXiv submitter | 50.08% | 17189 |
| I am an arXiv (other type of user): Please describe | 2.46% | 845 |

The number of articles I have published/submitted on arXiv is:

| Answer | % | Count |
|---|---|---|
| 1 article | 11.99% | 2570 |
| 2 articles | 8.96% | 1920 |
| 3 - 4 articles | 15.19% | 3254 |
| 5-10 articles | 23.06% | 4941 |
| More than 10 articles | 40.80% | 8743 |
| Total | 100% | 21428 |

As a user, my main subject area of interest in arXiv is: (please choose all that apply):

| Answer | % | Count |
| --- | --- | --- |
| Physics | 63.00% | 21193 |
| Mathematics | 33.19% | 11165 |
| Computer Science | 22.09% | 7430 |
| Other (please specify) | 7.37% | 2479 |
| Quantitative Biology | 3.68% | 1237 |
| Quantitative Finance | 2.21% | 745 |
| Statistics | 8.13% | 2734 |

Would you like to see additional subject categories added to arXiv?

| Answer | % | Count |
| --- | --- | --- |
| No | 73.76% | 22537 |
| Yes | 26.24% | 8019 |
| Total | 100% | 30556 |

I have been using arXiv for:

| Answer | % | Count |
| --- | --- | --- |
| 0 - 2 years | 19.54% | 6470 |
| 3 - 5 years | 28.96% | 9592 |
| 6- 10 years | 25.44% | 8425 |
| 11 or more years | 26.06% | 8632 |
| Total | 100% | 33119 |

My main place of work is:

| Answer | % | Count |
|---|---|---|
| United States of America | 26.45% | 8268 |
| Germany | 8.59% | 2686 |
| China | 7.37% | 2305 |
| United Kingdom of Great Britain and Northern Ireland | 5.54% | 1731 |
| France | 5.45% | 1703 |
| India | 4.40% | 1376 |
| Italy | 3.42% | 1069 |
| Japan | 3.37% | 1054 |
| Canada | 3.11% | 973 |
| Brazil | 2.35% | 734 |
| Russian Federation | 2.28% | 712 |
| Spain | 1.92% | 599 |
| Switzerland | 1.84% | 576 |
| Australia | 1.73% | 541 |
| Netherlands | 1.57% | 491 |
| Sweden | 1.02% | 320 |
| Israel | 1.01% | 317 |
| Poland | 0.98% | 307 |
| 109 other countries | 15.11% | 4736 |
| Total | 100% | 30498 |

My current occupation is: (Please choose all that apply)

| Answer | % | Count |
|---|---|---|
| I am an academic faculty member (professor) at a college or university | 26.98% | 8868 |
| I am an academic staff member (researcher or postdoc) at a college or university | 21.92% | 7207 |
| I am a researcher at a non-profit or governmental agency | 8.23% | 2707 |
| I am a Masters/Ph.D. student | 30.08% | 9890 |
| I am an undergraduate student | 4.61% | 1514 |
| I am (please describe) | 13% | 4353 |

My age is:

| Answer | % | Count |
|---|---|---|
| younger than 30 years | 37.42% | 12364 |
| 30 - 39 years | 31.27% | 10332 |
| 40 - 49 years | 13.76% | 4545 |
| 50 - 59 years | 9.30% | 3073 |
| 60 - 69 years | 5.77% | 1908 |
| 70 years and over | 2.47% | 817 |
| Total | 100% | 33039 |

**SECTION 2**

How important is it to improve on the following CURRENT arXiv services?

| Question | Very important | Important | Somewhat important | Not important | Should not be doing this | No opinion | Total |
|---|---|---|---|---|---|---|---|
| Simplify the submission process by providing clearer instructions and simpler language. | 3973 | 5048 | 6269 | 6419 | 585 | 5503 | 27797 |
| Improve support for submitting research papers by updating the TeX engine. | 4247 | 6641 | 6408 | 4330 | 292 | 5742 | 27660 |
| Improve support for submitting research data, code, slides, and other materials associated with a paper (e.g., I want to be able to upload my datasets/machine readable tables with my article). | 4529 | 7060 | 6255 | 3477 | 400 | 5903 | 27624 |
| Improve support for linking research data, code, slides, and other materials associated with a paper (e.g., I want to be able to link to my slides on SlideShare). | 4211 | 7121 | 7026 | 4343 | 591 | 4585 | 27877 |
| Improve search functions to allow more refined results (e.g., narrow down results by additional search terms, filter by publication year or institutional affiliation, etc.). | 10152 | 9796 | 5483 | 1631 | 110 | 1172 | 28344 |
| Improve the email alert system so that readers can customize their settings and choose to receive alerts about specific sub-topics. | 3969 | 6563 | 7370 | 5354 | 282 | 4291 | 27829 |
| Improve the trackback mechanism (linking papers back to blogs and commentaries that cite those papers). | 3342 | 6822 | 8211 | 4851 | 800 | 3805 | 27831 |

How important is it to develop the following NEW arXiv services?

| Question | Very important | Important | Some-what important | Not important | Should not be doing this | No opinion | Total |
|---|---|---|---|---|---|---|---|
| Enable submitting an article to a journal at the same time as it is uploaded to arXiv. | 3917 | 6536 | 6145 | 5294 | 1421 | 3299 | 26612 |
| Provide Citation Analysis tools (examining the frequency and pattern of a pattern of a paper's citation). | 5696 | 8542 | 7281 | 2985 | 855 | 1531 | 26890 |
| Offer citation export in formats such as BibTeX, RIS, etc. | 6939 | 8463 | 6141 | 2742 | 183 | 2235 | 26703 |
| Enable extraction for the BibTeX entry for the arXiv citation. | 6191 | 8468 | 6288 | 2427 | 139 | 2880 | 26393 |
| Add direct links to papers in the references (support reference extraction). | 6282 | 10589 | 7195 | 1430 | 118 | 1148 | 26762 |
| Support compliance with public/open access mandates (funding agency policies that require research results to be made public) by allowing final versions of papers to be submitted with information such as funding sources and grant numbers. | 3985 | 7081 | 6897 | 3314 | 285 | 4750 | 26312 |
| Enable linkages (interoperability) with other repositories (e.g., run by libraries), so that a paper accepted by arXiv is accepted at the same time by the other repositories. | 2809 | 6458 | 7397 | 4059 | 475 | 5091 | 26289 |
| Develop an annotation feature which will allow readers to comment on papers. | 3897 | 5457 | 6334 | 4127 | 5010 | 1987 | 26812 |
| Offer a rating system so readers can recommend arXiv papers that they find valuable. | 4179 | 5581 | 5854 | 3694 | 5873 | 1723 | 26904 |

Where do you go to find arXiv papers? Please choose all that apply.

| Answer | % | Count |
| --- | --- | --- |
| Go directly to arXiv.org (arXiv homepage) | 79% | 22804 |
| ADS | 14% | 4144 |
| Inspire | 13% | 3773 |
| Google Scholar | 35% | 10016 |
| Google search engine | 50% | 14440 |
| arXiv email alerts | 14% | 4086 |
| Other search engines | 5% | 1402 |
| Subject gateways for arXiv, such as the Math Front | 4% | 1203 |
| Other (please specify): | 9% | 2662 |

If you have used the arXiv homepage for finding papers, how easy is it to navigate?

| Answer | % | Count |
| --- | --- | --- |
| Very easy | 14.85% | 3916 |
| Easy | 32.05% | 8450 |
| Somewhat easy | 25.20% | 6644 |
| Somewhat difficult | 21.60% | 5696 |
| Difficult | 5.02% | 1324 |
| Very difficult | 1.27% | 336 |
| Total | 100% | 26366 |

If you have used the arXiv homepage, how do you usually navigate our main page? Please choose all that apply.

| Answer | % | Count |
|---|---|---|
| Go to link "new" or "recent" under a particular category | 63% | 16503 |
| Use arXiv search engine and enter a specific arXiv-id, author name, or search term | 63% | 16478 |
| Receive daily mailing list, and then look for a particular article on the search field | 14% | 3692 |
| Other, please explain: | 3% | 853 |

## SECTION 3

How important are the following CURRENT quality control measures? (Very important/Important/Somewhat important/Not important and should not be doing this/No opinion)

| Question | Very important | Important | Somewhat important | Not important | Should not be doing this | No opinion | Total |
|---|---|---|---|---|---|---|---|
| arXiv keeps out (rejects) papers that don't have much scientific value. | 7935 | 7928 | 5057 | 2066 | 2029 | 1413 | 26428 |
| arXiv checks papers for text overlap: an author's use of too much identical text from other authors' papers, without making it clear that the text is not their own material, i.e., "plagiarism". | 11109 | 9336 | 3873 | 934 | 376 | 783 | 26411 |
| arXiv checks papers for too much text re-use from an author's earlier works, i.e., "self-plagiarism" (reuse of identical content from one's own published work without citing). | 6411 | 8783 | 6481 | 2681 | 1023 | 924 | 26303 |
| arXiv checks papers for format-related problems (line numbers in text, missing references, oversize submissions, etc.) and asks authors to fix them before they are announced. | 4424 | 9955 | 7799 | 2446 | 563 | 958 | 26145 |
| arXiv makes sure submissions are correctly classified (the subject categories are included on the arXiv homepage). | 5868 | 10991 | 6630 | 1483 | 354 | 862 | 26188 |
| arXiv moderates the scientific content of trackback (links to blogs and commentaries) before permitting the link to be added. | 3137 | 7082 | 6790 | 3414 | 1126 | 4258 | 25807 |

Please choose any ONE of the following statements that you agree with the most:

| Answer | % | Count |
|---|---|---|
| arXiv should not perform quality control at all; arXiv should focus on quickly publishing papers. | 4.80% | 1289 |
| arXiv should focus on a few parts of quality control, but mostly focus on quickly publishing papers. | 31.42% | 8432 |
| arXiv needs to keep up quality control in many areas and, if needed, should take a little more time to publish papers so that quality control can be kept. | 17.24% | 4626 |
| arXiv should quickly publish and at the same time, keep up with quality control, even if that takes time away from other new activities. | 9.76% | 2620 |

## SECTION 4

Which of the following BEST describes your opinion of how arXiv needs to move forward?

| Answer | % | Count |
|---|---|---|
| arXiv should focus on its main purpose, which is to quickly make available scientific papers. This will be enough to hold up the value of arXiv in the future. | 71.94% | 19865 |
| arXiv should expand its main mission, and spend more time and resources to provide new services. This is necessary to hold up the value of arXiv in the future. | 19.59% | 5410 |
| No opinion | 8.47% | 2340 |
| Total | 100% | 27615 |

Overall, how satisfied are you with arXiv?

| Answer | % | Count |
|---|---|---|
| Very satisfied | 52.92% | 14770 |
| Satisfied | 42.43% | 11841 |
| Somewhat satisfied | 3.55% | 990 |
| Somewhat dissatisfied | 0.54% | 150 |
| Very dissatisfied | 0.15% | 42 |
| No opinion | 0.42% | 116 |
| Total | 100% | 27909 |